\documentclass[aps,prl,twocolumn,showpacs,groupedaddress,floatfix]{revtex4}

\usepackage{graphicx} 
\usepackage{dcolumn}
\usepackage{epsfig}
\usepackage{amssymb} 
\usepackage{amsmath} 
\usepackage{rotating} 
\usepackage{color} 
\newcommand{\gammap}{\dot{\gamma}}
\newcommand{\seuil}{\sigma_{\scriptscriptstyle 0}}

\begin{document}

\title{Velocity profiles in shear-banding wormlike micelles}

\author{Jean-Baptiste Salmon}
\affiliation{Centre de Recherche Paul Pascal, Avenue Schweitzer, 33600 Pessac, FRANCE}
\author{Annie Colin}
\affiliation{Centre de Recherche Paul Pascal, Avenue Schweitzer, 33600 Pessac, FRANCE}
\author{S\'ebastien Manneville}
\affiliation{Centre de Recherche Paul Pascal, Avenue Schweitzer, 33600 Pessac, FRANCE} 
\author{Fran\c{c}ois Molino}
\affiliation{Groupe de Dynamique des Phases Condens\'ees, Universit\'e Montpellier II, Place E. Bataillon, 34095 Montpellier, FRANCE}
\date{\today}
\begin{abstract}
Using Dynamic Light Scattering in heterodyne mode, we measure velocity profiles
 in a much studied system of wormlike micelles (CPCl/NaSal) known to exhibit
 both shear-banding and stress plateau behavior. Our data provide evidence
 for the simplest shear-banding scenario, according to which the
 effective viscosity drop in the system is due to the nucleation and growth
 of a highly sheared band in the gap, whose thickness linearly increases with the
 imposed shear rate. We discuss various details of the velocity profiles in
 all the regions of the flow curve and emphasize on the complex, non-Newtonian nature of
 the flow in the highly sheared band. 
\end{abstract}
\maketitle

Understanding the  correlation between mechanical and structural
 response in non-Newtonian fluids submitted to  high deformation
 rates is crucial on both fondamental and  technological grounds \cite{Larson:99}.
 Among the  variety of complex fluids investigated in recent years,
 a wide class exhibits flow--structure coupling that leads to a strong
 shear-thinning behavior: along the steady-state flow curve (shear stress $\sigma$
 vs. shear rate $\gammap$),
 a drop of up to three orders of magnitude in the effective viscosity $\eta=\sigma/\gammap$
 is observed in a very narrow stress range
 leading to a stress plateau (for a review, see for instance
 Refs.~\cite{Larson:99,Edimbourg:00}). In correlation with
 this stress plateau, bands of different micro-structures and normal
 to the velocity gradient appear. Such bands correspond to a new shear-induced
 structure (SIS), whose low viscosity is in general supposed to be
 responsible for the shear-thinning. This so-called {\it shear-banding} behavior has
 been observed in both ordered mesophases (lamellar, hexagonal, cubic)
 \cite{Roux:93}
 and transient gels \cite{Molino:00}.

\begin{figure}[htbp]
\begin{center}
\scalebox{0.65}{\includegraphics{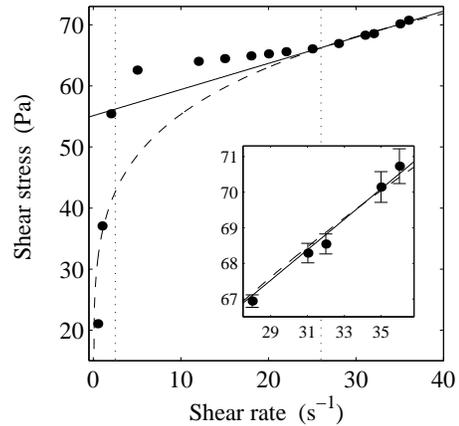}}
\end{center}
\caption{\label{rheol} Steady-state flow curve
 for a 6~\% wt. CPCl/NaSal
 solution in 0.5~M brine at 21.5$^\circ$C. The data were collected under
 controlled shear rate in a 1~mm gap Couette cell. Vertical lines
 indicate the limits of the banding domain. For
 $\gammap>\gammap_{\scriptscriptstyle 2}\approx 26$~s$^{-1}$, the data were fitted
 by a power-law fluid $\sigma=35.9\,\gammap^{0.19}$ (dashed line)
 and by a Bingham fluid
 $\sigma=55.1+0.43\,\gammap$ (solid line). Inset:
 high shear branch; error bars account for temporal fluctuations of $\sigma$.
}
\end{figure}

A particularly
 well-documented example is the group of wormlike micellar systems
 of self-assembled surfactant molecules \cite{Berret:94,Schmitt:94}. They consist of very long
 cylindrical aggregates whose configurations mimic polymer solutions.
 However their dynamics is strongly modified by the equilibrium character
 of the chains, which enables them to break and recombine \cite{Cates:87}. Generically,
 one starts from an isotropic viscoelastic solution of these micelles
 above the semidilute regime, which behaves like a Maxwell fluid at
 low shear rates. Upon increasing $\gammap$ and entering the nonlinear
 regime, the onset of the stress plateau for a critical shear rate $\gammap_{\scriptscriptstyle 1}$
 is associated with the nucleation and growth of highly birefringent
 bands, suggesting strong alignment of the micelles along the velocity
 direction \cite{Berret:94,Schmitt:94}.
 As the shear rate is further increased above $\gammap_{\scriptscriptstyle 1}$,
 the new organization progressively fills the gap at almost constant
 stress, up to a second critical shear rate $\gammap_{\scriptscriptstyle 2}$.
 Above $\gammap_{\scriptscriptstyle 2}$, the system
 enters a second regime of apparently homogeneous structure, with
 a second branch of increasing stress. The flow curve of Fig.~\ref{rheol}
 is typical of micellar systems like that investigated in the present work.
 Such a stress plateau has been reported for concentrations close to the
 equilibrium isotropic--nematic (I--N) transition, where  coupling between the
 order parameter and the flow could explain an out-of-equilibrium phase
 transition to local nematic order \cite{Schmitt:94,Olmsted:97}. 
This behavior has also been observed in more dilute systems,
 where arguments based on instabilities in the underlying 
flow curve have been invoked \cite{Berret:94,Spenley:93}.

In the earliest simple pictures
 of the flow along the stress plateau,
 the system is supposed to separate into two differently sheared bands:
 a weakly sheared region that flows at $\gammap_{\scriptscriptstyle 1}$
 and a highly sheared region at $\gammap_{\scriptscriptstyle 2}$
 \cite{Olmsted:97,Spenley:93}.
 Experimentally, this very basic issue of the nature of the flow field
 remains rather obscure.
 Among previous studies, Nuclear Magnetic Resonance
 (NMR) imaging has proved the most
 successful technique to measure local velocity and/or local shear rate.
 Callaghan and coworkers have
 shown for the first time the existence of inhomogeneous velocity fields
 in wormlike micelles \cite{Mair:96,Britton:97,Fischer:01}.
 However, due to the limited amount of data in NRM images, no clear evidence 
 supporting the simple shear-banding scenario has been provided. In particular, one of
 the most detailed NMR studies promotes a quite different dynamical description:
 instead of being submitted to a high shear rate, the nucleated nematic-like structure
 seems to undergo a zero-shear-rate plug-like
 flow and to move like a piece of gel, probably involving important
 wall slip and fracture behaviors \cite{Fischer:01}. 
 This last observation has put the
 whole naive shear-thinning picture on uncertain ground.

In this Letter, we show that for the much studied wormlike
 micellar system CPCl/NaSal in brine and for concentrations
 far from the I--N transition, the
 simplest scenario holds. The originality of our work
 relies on recording both the local velocity
 and global rheological data in Couette geometry
 simultaneously along the whole flow curve.
 This enables to demonstrate for the first time
 the nucleation of a highly sheared band at
 a critical stress and to follow its growth from the rotor to the stator
 as the shear rate spans the stress plateau.
 We show that the width of the band grows linearly
 with the shear rate, in agreement with the decrease
 of the measured effective viscosity, and that the SIS
 is not Newtonian.

Our system of elongated wormlike micelles consists of a binary
 mixture of cetylpyridinium chloride (CP$^+$, Cl$^-$) and sodium
 salicylate (Na$^+$, Sal$^-$) in 0.5~M NaCl--brine. Since the
 pioneering work by Hoffmann and
 Rehage \cite{Rehage:91}, this extensively studied system
 has been demonstrated to exhibit a stress plateau and
 optical birefringence shear-banding \cite{Berret:94}. We focus on
 a 6~\% wt. sample at 21.5$^{\circ}$C, in the domain
 above the semidilute regime (0.5--5~\% wt.) but far below
 the equilibrium I--N transition (around 20~\% wt.).
 For the concentrations considered here, the system is known to be a perfect
 Maxwell fluid in the linear regime, with a typical relaxation time
 of about 1~s.

Rheological flow curves are measured using a standard rheometer
 (TA Instruments AR 1000) and transparent Couette cells of
 outer radius 25~mm and different
 gap widths $e$ ($e=1$ or 3~mm). The temperature is
 maintained at $21.5\pm 0.1^{\circ}$C by
 a water circulation around the cell.
 To access local velocity, we use a heterodyne
 Dynamic Light Scattering (DLS) technique that has been described
 elsewhere \cite{Salmon:02_2}. 
 The measurement of the local
 velocity relies on performing the
 interference between a reference beam and light scattered
 from a small volume of the sample of typical size 50~$\mu$m. The
 correlation function of the interference signal
 exhibits oscillations at the Doppler frequency $\mathbf{q}\cdot\mathbf{v}$
 where  $\mathbf{q}$ is the scattering wavevector and $\mathbf{v}$
 the local velocity. Good statistical convergence is achieved 
 by averaging the correlation function over 3~s.
 Velocity profiles are measured by moving the
 rheometer along the velocity gradient
 by steps of 30~$\mu$m. This technique
 enables us to obtain a complete velocity profile
 in about 2~min, more than 10 times faster than NMR velocity imaging.
 Typical uncertainties are about 5~\%.

In order to enhance the scattering properties of our
 system, we add a small amount (1--5~\% wt.) of 30~nm colloidal particles
 (Ludox from Aldrich).
 We checked that the rheological properties, particularly
 the plateau behavior, was not affected by the addition of those scatterers.
 Indeed, the small size of the scatterers compared
 to the typical mesh size of the micellar network in this concentration 
 domain should lead to a negligible influence on the structural and
 mechanical behaviors of the sample.

\begin{figure}[htbp]
\begin{center}
\scalebox{0.5}{\includegraphics{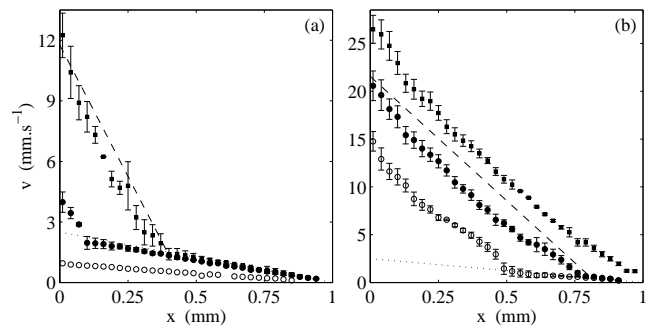}}
\end{center}
\caption{\label{profils} Velocity profiles recorded
 while measuring the flow curve of Fig.~\ref{rheol}.
 (a) $\gammap=1~(\circ), 5~(\bullet),$ and 12~s$^{-1}~({\scriptscriptstyle\blacksquare})$.
 (b) $\gammap=15~(\circ), 22~(\bullet),$
 and 28~s$^{-1}~({\scriptscriptstyle\blacksquare})$. The value $x=0$ ($x=1$ resp.)
 corresponds to the rotor (stator resp.).
 The dotted line is $v(x)=\gammap_{\scriptscriptstyle 1}\,(e-x)$ with
 $\gammap_{\scriptscriptstyle 1}=2.5$~s$^{-1}$. The dashed lines correspond to a
 shear rate $\gammap_{\scriptscriptstyle 2}=26$~s$^{-1}$}
\end{figure}

Figure~\ref{rheol} shows the steady-state shear stress
 recorded in our micellar system at various imposed shear rates.
 This flow curve presents a stress plateau at
 $\sigma\approx 65$~Pa
 that extends from $\gammap_{\scriptscriptstyle 1}\approx 2.5$~s$^{-1}$ to $\gammap_{\scriptscriptstyle 2}\approx 26$~s$^{-1}$
 corresponding to a drop in the effective viscosity by a factor of 10. Note that for
 $\gammap>\gammap_{\scriptscriptstyle 2}$, the stress response is no longer strictly
 stationary: $\sigma$ fluctuates by about 2~\% around its mean value. Finally, at
 $\gammap\approx 37$~s$^{-1}$, the sample began to fracture so that no
 measurement is available at higher shear rates.
 
\begin{figure}[htbp]
\begin{center}
\scalebox{0.65}{\includegraphics{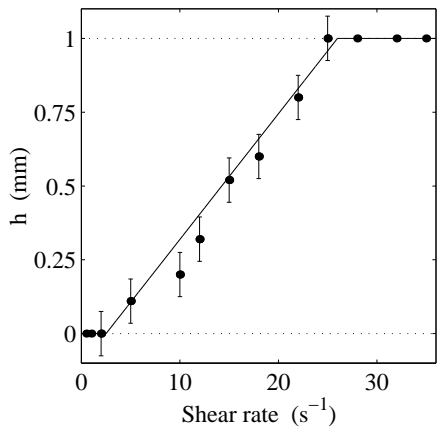}}
\end{center}
\caption{\label{bande} Width $h$ of the highly sheared
 band as a function of $\gammap$.
 With the value $\gammap_{\scriptscriptstyle 1}=2.5\pm 0.2$~s$^{-1}$ inferred from
 the velocity profiles of Fig.~\ref{profils}, a good linear
 approximation of $h(\gammap)$ is obtained with
 $\gammap_{\scriptscriptstyle 2}=26\pm 1$~s$^{-1}$ (solid line).
}
\end{figure}

Figure~\ref{profils} summarizes our main result:
 velocity profiles in the gap are displayed
 for $\gammap$ ranging from 1~s$^{-1}$ to 28~s$^{-1}$.
 For all the shear rates in the plateau domain, the velocity profiles
 exhibit an unambiguous banding structure: two regions of
 different well-defined local shear rates
 coexist when $\gammap_{\scriptscriptstyle 1}<\gammap<\gammap_{\scriptscriptstyle 2}$.
 The highly sheared band
 nucleates on the inner wall of the Couette cell at the onset of
 the plateau and expands
 as $\gammap$ is increased up to $\gammap_{\scriptscriptstyle 2}$ where the band fills the whole gap.
 As can be seen on Fig.~\ref{profils},
 the local shear rate in the weakly sheared region remains
 constant and equal to $\gammap_{\scriptscriptstyle 1}=2.5\pm0.2$~s$^{-1}$.
 Moreover, although significantly curved (see below),
 the velocity profiles in the highly sheared region are compatible
 with a local shear rate of $\gammap_{\scriptscriptstyle 2}=26\pm 1$~s$^{-1}$.
 Note that in all our data, no noticeable wall slip was observed.
 Finally, these profiles easily yield the width $h$ of the highly sheared
 band. Figure~\ref{bande} demonstrates that $h$ increases linearly
 with $\gammap$. This corresponds to the simplest
 relation between the imposed shear rate and
 the two local values $\gammap_{\scriptscriptstyle 1}$ and $\gammap_{\scriptscriptstyle 2}$:
 $\gammap=(1-\alpha)\gammap_{\scriptscriptstyle 1}+\alpha\gammap_{\scriptscriptstyle 2}$,
 where $\alpha=h/e$ increases from 0 to 1 over the plateau region.

\begin{figure}[htbp]
\begin{center}
\scalebox{0.65}{\includegraphics{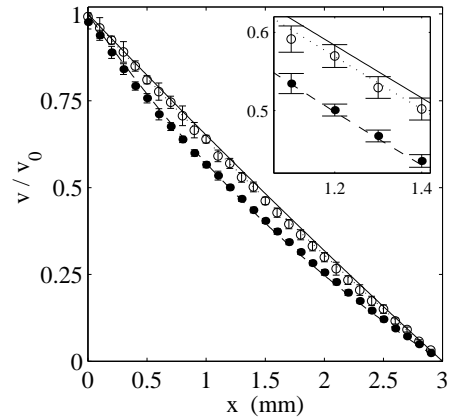}}
\end{center}
\caption{\label{profil} Velocity
 profiles measured in the 3~mm gap Couette
 cell and rescaled by the rotor velocity $v_{\scriptscriptstyle 0}$
 for two values of $\gammap$ below the banding
 transition.
 For $\gammap=1$~s$^{-1}~(\circ)$, the normalized data are
 well fitted by a weakly shear-thinning law
 $\sigma\sim\gammap^{0.7\pm 0.1}$ (dotted line).
 For $\gammap=2$~s$^{-1}~(\bullet)$, a stronger shear-thinning
 behavior $\sigma\sim\gammap^{0.28\pm 0.03}$ provides a
 very good fit (dashed line).
 The solid line is the velocity profile expected for a Newtonian fluid.
}
\end{figure}

Beyond shear-banding evidence, the resolution
 of our technique enables us to analyze in more
 detail the flow behavior of the material in
 the regions below and above the plateau. In
 Fig.~\ref{profil} are displayed
 two normalized velocity profiles below the plateau
 at $\gammap=1$~s$^{-1}$ and $\gammap=2$~s$^{-1}$
 in a 3~mm gap. For $\gammap=1$~s$^{-1}$,
 the velocity profile is very close to a straight line,
 consistent with the Newtonian behavior of the micellar
 solution at low shear rates. However, as shown in the inset
 of Fig.~\ref{profil}, our data do not exactly fall on
 the Newtonian velocity profile but rather present a small
 curvature which can be accounted for by a power-law
 $\sigma\sim\gammap^{0.7\pm 0.1}$.
 When $\gammap=2$~s$^{-1}$ {\it i.e.} just below the onset
 of shear-banding at $\gammap_{\scriptscriptstyle 1}$, the curvature is much more pronounced and
 the power law $\sigma\sim\gammap^{0.28\pm 0.03}$ yields
 a perfect fit of the data. This demonstrates
 the existence of weak shear-thinning on the low shear branch,
 that sharply increases as the banding transition is approached.

\begin{figure}[htbp]
\begin{center}
\scalebox{0.5}{\includegraphics{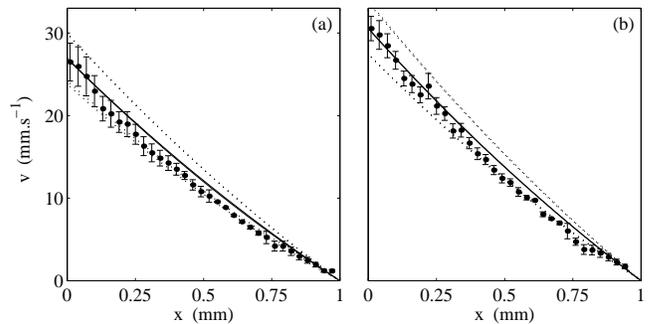}}
\end{center}
\caption{\label{power} Velocity profiles in
 the high shear branch at (a) $\gammap=28$~s$^{-1}$ and
 $\sigma=66.9$~Pa and
 (b) $\gammap=32$~s$^{-1}$ and
 $\sigma=68.5$~Pa. In each case, two solid lines correspond to two different
 but undistinguishable fits: (i)
 a power-law behavior $\sigma=35.9\,\gammap^{0.19}$ and (ii)
 a Bingham behavior $\sigma=55.1+0.43\,\gammap$.
 The dotted lines account for a global stress
 fluctuation of 2~\%. 
}
\end{figure}
 
Looking more closely at the velocity profiles in the plateau region,
 one notes that
 the highly sheared band displays a significant curvature
 even in a 1~mm gap (see Fig.~\ref{profils}(b)). This shows
 that the SIS cannot be described by a Newtonian fluid.
 The same observation holds for velocity profiles
 recorded at $\gammap>\gammap_{\scriptscriptstyle 2}$ where the SIS
 has completely invaded the gap.
 A precise analysis of those velocity profiles may help to understand
 both the global rheological measurements and the behavior of the
 SIS.
 Indeed, if one attempts to fit the flow curve of
 Fig.~\ref{rheol} for $\gammap>\gammap_{\scriptscriptstyle 2}$ according
 to a power-law $\sigma=A\,\gammap^n$, the result is dramatically
 shear-thinning: $\sigma=35.9\,\gammap^{0.19}$ (see inset of Fig.~\ref{rheol}).
 Such a shear-thinning exponent $n$
 allows us to account for the curvature of
 all the velocity profiles at $\gammap>\gammap_{\scriptscriptstyle 2}$ with a single prefactor
 $A$ consistent with the rheological data
 (see Fig.~\ref{power} for the cases  $\gammap=28$ and 32~s$^{-1}$
 where the stress has been allowed to vary by 2~\% according to the
 measured temporal fluctuations of $\sigma$).

However, above $\gammap_{\scriptscriptstyle 2}$,
 other constitutive
 relations $\sigma(\gammap)$ may
 fit the experimental data as well. For instance, a Bingham
 fluid $\sigma=\seuil+A\gammap$ with $\seuil=55.1$~Pa and
 $A=0.43$~Pa.s also closely fits both the flow curve
 and velocity profiles for $\gammap>\gammap_{\scriptscriptstyle 2}$ with a single set of
 parameters $(\seuil,A)$ (see inset of Fig.~\ref{rheol} and Fig.~\ref{power}).
 Such a Bingham behavior is suggested from the analogy with
 other related viscoelastic systems,
 whose similar nonlinear behavior has been
 interpreted in terms of a steady-state system of bulk fractures \cite{Molino:00}.
 Here, the range of $\gammap$ on the high shear branch
 is limited to 26--37~s$^{-1}$ so that we cannot discriminate between
 the two above behaviors and
 only further studies will help
 to select the correct non-Newtonian
 relation for the high shear branch. 

Let us now further discuss our results in light of previous studies.
 Among the great diversity of wormlike micellar systems, a universal feature
 seems to emerge: the presence of a robust stress plateau associated
 to the growth of a SIS. However the present results
 and recent local NMR velocity measurements by Fischer {\it et al.} \cite{Fischer:01} lead
 to two radically different descriptions of the flow field. Ref.~\cite{Fischer:01}
 suggests the nucleation and growth of a gel whereas our data clearly shows
 the coexistence of two different well-defined shear bands flowing at
 $\gammap_{\scriptscriptstyle 1}$ and $\gammap_{\scriptscriptstyle 2}$
 respectively.

These important discrepancies may be explained by two
 different arguments. First, the distance to the I--N transition is
 drastically different in the systems under study. Indeed,
 Fischer {\it et al.} used
 a CTAB/D$_{\scriptscriptstyle 2}$O system at a surfactant concentration
 of 20~\% wt. very close to the I--N transition ($\approx 21~\%$ wt. at the
 working temperature). In our case (CPCl/NaSal in brine), the surfactant
 concentration of 6~\% wt. was much more distant from the I--N transition.
 Second, significant temporal fluctuations of the flow field are
 mentioned in both the present work and Refs.~\cite{Britton:97,Fischer:01}.
 If the flow of the SIS is time-dependent, we think that
 the data analysis in the two techniques may be affected in different ways.
 The temporal resolution of our heterodyne
 DLS setup provides an accurate value of the local velocity
 averaged over a few seconds. With NMR imaging, a broad velocity
 distribution is reported in the SIS, sometimes showing several
 peaks. We believe that this rich spectral content may 
 complicate the interpretation of the average NMR data.
 
Another important open issue concerns the very nature
of the SIS and its rheological behavior. 
As far as the structure is concerned, 
birefringence appears as a robust experimental
fact \cite{Berret:94,Schmitt:94}. 
However, the correlation between birefringence
and local nematic order is only confirmed for
systems close to the I--N transition \cite{Britton:97}.
This points to possible differences between shear-induced structures
far from and close to this equilibrium transition.  
As for the behavior under shear, flow instabilities in the SIS at high 
shear rates leads to fracture and ejection of the sample from the rheometer
\cite{Berret:94,Spenley:93}. 
We believe that the non-Newtonian features of the SIS revealed in this
Letter may play an important role in such instabilities.
Indeed temporal fluctuations of the stress and of the birefringent
bands seem to suggest intermittent and localized
fracture-like flow events, yet not directly observed so far.

 Although separate studies of rheology, birefringence,
 and more recently NMR velocity profiles, have been performed,
 a complete picture of the nonlinear rheology of wormlike
 micelles remained rather elusive in the literature due to the lack of
 experimental data. Coupled with
 optical birefringence and rheological measurements, our heterodyne DLS
 setup appears as a well-suited tool to provide the missing information 
 on this much studied problem.

\begin{acknowledgments}
The authors are grateful to the 
 ``Cellule Instrumentation'' at CRPP for
 building the experimental setup.
 We also thank J.-F.~Berret, C.~Gay, and P.~Olmsted
 for fruitful discussions. 
\end{acknowledgments}


\end{document}